\centerline {\bf A NEW
REALIZATION OF HOLOGRAPHY } \bigskip \centerline { Carlos Castro }
\bigskip \centerline { Center for Theoretical Studies of Physical Systems
} \centerline { Clark Atlanta University, Atlanta, GA. 30314 } \bigskip

\centerline{\bf Abstract} \bigskip Based on the old results of Cho, Soh,
Park and Yoon, it is shown how higher $m+ n $ dimensional pure
gravitational actions restricted to $ AdS_m \times S^n $ backgrounds admit
a holographic reduction to a lower $ m$-dimensional Yang-Mills-like gauge
theory of diffeomorphisms of $ S^n $ interacting with a charged non-linear
sigma model plus boundary terms by a simple tuning of the $ AdS_m$-throat
and $S^n$-radius sizes. After performing a harmonic expansion of the
fields with respect to the internal coordinates and a subsequent
integration one obtains an $m$-dimensional effective action involving an
infinite-component field theory.  The supersymmetrization program can be
carried out in a straighforward fashion. \bigskip In the past years there
has been a large activity on the holographic principle and the AdS/CFT
correspondence [1,2]. It is unfortunate that the work of [3] received
little attention when in fact contained already the seeds of the
holographic principle. The main result of [3] is that $m+n$-dim Einstein
gravity can be identified with an $m$-dimensional generally invariant
gauge theory of Diffs $N$, where $N$ is an $n$-dim manifold. Locally the
$m+n$-dim space can be written as $ \Sigma = M \times N $ and the metric
$g_{AB}$ decomposes: $$ g_{AB} \Rightarrow \gamma_{\mu\nu} ( x, y ) + e^2
\phi_{ab} ( x, y ) A^a_\mu ( x, y ) A^b_{\nu} ( x, y ) . ~~~ e A^a_\mu (
x, y ) \phi_{ab} ( x, y ). ~~~ \phi_{ab} ( x, y ) .... \eqno (1) $$ where
$e$ is the gauge coupling constant. This decomposition must $not$ be
confused with the Kaluza-Klein reduction where one imposes an isometry
restriction on the $g_{AB}$ that turns $A^a_\mu$ into a gauge connection
associated with the gauge group $G$ generated by isometry. Dropping the
isometry restrictions allows $all$ the fields to depend on $all$ the
coordinates $x, y $. Nevertheless $ A^a_\mu ( x, y )$ can still be
identified as a connection associated with the infinite-dim gauge group of
Diffs $N$. The gauge transformations are now given in terms of
Lie-brackets and Lie derivatives: $$ \delta A^a_\mu = - { 1\over e } D_\mu
\xi^a = - { 1\over e } ( \partial_\mu \xi^a - e [ A_\mu, \xi ]^a ) = - { 1
\over e } ( \partial_\mu - e {\cal L}_{A_\mu } ) \xi^a $$ $$ A_\mu \equiv
A^a_\mu \partial_a. ~~~ {\cal L}_{ A_\mu } \xi^a \equiv [ A_\mu, \xi ]^a
$$ $$\delta \phi_{ab} = - [ \xi , \phi ]_{ab} . ~~~ \delta \gamma_{\mu\nu}
= - [ \xi, \gamma_{\mu\nu } ] . \eqno (2) $$ In particular, if the
relevant algebra is the area-preserving diffs of $S^2 $, given by the
suitable basis dependent limit of the large $ N$ limit, $ SU(\infty)$ [4],
one induces a natural Lie-Poisson structure generated by the gauge fields
$ A_\mu $. The Lie derivative of $f$ along a vector $\xi$ is the Lie
bracket $ [ \xi, f ] $ which coincides in this case with the Poisson
bracket $\{ \xi, f \}$. This implies that the Lie brackets of two
generators of the area-preserving diffs $S^2$ is given precisely by the
generator associated with their respective Poisson brackets (a Lie-Poisson
structure): $$ [ L_f , L_g ] = L _{ \{f, g \} }. \eqno ( 3) $$ This
relation is derived by taking the vectors $\xi_1^a, \xi_2^a$, along which
we compute the Lie derivatives, to be the symplectic gradients of two
functions $ f(\sigma^1, \sigma^2), g(\sigma^1, \sigma^2 ) $: $$ \xi^a_1 =
\Omega^{ab}\partial_b f. ~~~ \xi^a_2 = \Omega^{ab}\partial_b g . \eqno ( 4
) $$ When nontrivial topologies are involved one must include harmonic
forms $\omega$ into the definition of $\xi^a$ allowing central terms for
the algebras. This relation can be extended to the volume-preserving diffs
of $N$ by means of the Nambu-Poisson brackets: $$ \{ A_1, A_2,
A_3,......A_n \} = Jacobian = { \partial ( A_1, A_2, A_3,....., A_n )
\over \partial ( \sigma^1, \sigma^2, ....\sigma^n ) } \Rightarrow $$ $$[
L_{ A_{1} } , L_{A_{2} }, .........., L_{A_{n} } ] = L_{ \{ A_{1} ,
A_{2},........., A_{n} \} } . \eqno ( 5 ) $$ which states that the
Nambu-commutator of $n$- generators of the volume-preserving diffs of
${\cal N} $ is given by the generator associated with their corresponding
Nambu-Poisson brackets. The generators are obtained in this case by taking
the multi-symplectic gradients of functions $f_1, f_2....f_{n-1} $ of
$\sigma^1, \sigma^2....\sigma^n$ given in terms of the inverse of the
multi-symplectic $n$-form $\Omega$: $$\xi_{(i)}^{a_1} = \Omega^{a_1 a_2
a_3....a_n} \partial_{a_2}f^{(i)}_1
\partial_{a_3}f^{(i)}_2....\partial_{a_n}f^{ (i) }_{n-1} .....\eqno ( 6 )
$$ When the dimension of ${\cal N} $ is even, locally one can write the
volume form in terms of products of area-forms $\Omega = \omega \wedge
\omega....$ ; i.e in an appropriate frame the Jacobian (Nambu-Poisson
bracket) factorizes locally into a product of ordinary Poisson brackets.
However this doesn't mean that area-preserving has a one-to-one
correspondence with volume preservings. There are volume-preserving diffs
that do not necessarily preserve areas. The suitable star product will be
the Zariski product related to the deformation program of Nambu-Poisson
mechanics [20] ; i.e. deformation theory in multi-symplectic geometry. The
curvature scalar $R^{ ( m+n ) } $ decomposes into [3]: $$ \sqrt {det
~G_{AB} } R^{ ( m+n) } \Rightarrow \sqrt{ - \gamma } ~\sqrt { \phi }
\times \{ ~ \gamma^{\mu\nu} {\tilde R}^{(m)}_{\mu\nu } + { e^2 \over 4 }
\phi_{ab} F^a_{\mu\nu} F^{b}_{\rho\tau } \gamma^{\mu\rho} \gamma^{\nu\tau
} + $$ $$ \phi^{ab}R^{ (n)}_{ab} + { 1\over 4} \gamma^{\mu\nu} \phi^{ab}
\phi^{cd} [D_\mu \phi_{ac} D_\nu \phi_{bd } - D_\mu \phi_{ab} D_\nu
\phi_{cd }] + $$ $${ 1\over 4} \phi^{ab} \gamma^{\mu\nu} \gamma^{\rho\tau}
[\partial_a \gamma_{\mu\rho} \partial_b \gamma_{\nu \tau } - \partial_a
\gamma_{\mu\nu} \partial_b \gamma_{\rho\tau } ] \} + $$ $$ [ \partial_\mu
( \sqrt { - \gamma} \sqrt {\phi } j^\mu ) - \partial_a ( \sqrt { - \gamma
} \sqrt { \phi } e A^a_\mu j^\mu ) + \partial_a ( \sqrt { - \gamma}
\sqrt{\phi } j^a ) ] \eqno ( 7) $$ The currents are [3]: $$ j^\mu =
\gamma^{\mu \nu } \phi^{ab}D_\nu \phi_{ab} . ~~~ j^a = \phi^{ab}
\gamma^{\mu \nu } \partial_b \gamma_{\mu \nu } . \eqno ( 8 ) $$ Therefore,
Einstein gravity in $m+n$-dim describes an $m$-dim generally invariant
field theory under the gauge transformations or $Diffs~N$. Notice how
$A^a_\mu$ couples to the graviton $\gamma_{\mu\nu} $, meaning that the
graviton is charged (gauged) in this theory and to the $\phi_{ab}$ fields.
The ``Ricci'' tensor of the horizontal space is a $gauged$ Ricci tensor
meaning that it is constructed using the derivatives $ \partial_\mu - e
A^a_\mu \partial_a $. The ``metric'' $\phi_{ab}$ on $N$ can be identified
as a $gauged $ non-linear sigma field whose self interaction potential
term is given by: $\phi^{ab}R_{ab}^{ (n)}$. The contribution of the
currents to the action is essential when there are boundaries involved;
i.e. the proyective/conformal boundary of $AdS $spaces.

When the internal manifold is a homogeneous space one can perform a
harmonic expansion of the fields w.r.t the internal $y$ coordinates, and
after integrating w.r.t these coordinates $y$, one will generate an
infinite-component field theory on the $m$-dimensional space., i.e. the
effecitive action in the horizontal space will involve a sum over an
infinity of field components [22]. This resembles the higher spin field
theories of [6,7]. Whereas a reduction of the $Diffs~ N$, via the inner
automorphims of a subgroup $G$ of the $Diffs~ N$, yields the usual
Einstein-Yang-Mills theory interacting with a nonlinear sigma field. But
in general, the theory described in [3] is by far richer than the latter
theory. A crucial fact of the decomposition (7) is that $each$ single term
in (7) is by itself independently invariant under $Diffs~ N$.

To prove the main point of this short note, let us assume that $ m = n = 5
$. For the particular case (i) that the horizontal space metric $
\gamma_{\mu\nu} $ is $uncharged$ ; i.e. independent of the $ y$
coordinates and equal to the $AdS_5$ metric and (ii) $ \phi_{ab} $ depends
solely on the internal $ y $ coordinates ; i.e. constant field
configurations (from the point of view of the horizontal manifold) for the
non-linear sigma model. But the gauge field still retains the $full$
dependence on $all$ the coordinates $x, y $. In this particular case the
action simplifies drastically for the following reasons:

When $ \gamma_{\mu \nu } ( x ) = g_{\mu \nu } ( AdS_5 )$ the $gauged$
Ricci scalar curvature coincides precisely with the negative-valued scalar
curvature $ R ( AdS_5 ) $. The ``Riccci'' scalar for the internal manifold
will coincide with the positive-valued Ricci scalar $ R ( S^5 ) $ and the
terms $ \partial_a \gamma_{\mu \nu }, .... $ in the action are zero. If
one takes the size of the $AdS_5$ throat to coincide with the value of the
internal radius of $ S^5 $ there will be an exact $cancellation$ of the
scalar curvatures and the $ 10$-dim gravitational action reduces to a $
5$-dim Yang-Mills-like gauge theory of Diffs ~ $ S^5 $ interacting with a
charged non-linear sigma model $ \phi_{ab} $ plus boundary terms: $$ - { 1
\over 16 \pi G}\sqrt { det ~ G } R^{ ( m+n) } \Rightarrow - { 1 \over 16
\pi G} \sqrt{ - \gamma } ~\sqrt { \phi } \times \{ { e^2 \over 4 }
\phi_{ab} F^a_{\mu\nu} F^{b}_{\rho\tau } \gamma^{\mu\rho} \gamma^{\nu\tau
} + $$ $$ { 1\over 4} e^2 \gamma^{\mu\nu} \phi^{ab} \phi^{cd} [ A_\mu,
\phi]_{ac} [A_\nu , \phi]_{bd } - ...... \} +.... boundary~ terms \eqno (
9) $$ since when $ \phi_{ab}$ depend on $y$ only, the covariant
derivatives become: $$D_\mu \phi_{ab} = [e A_\mu, \phi]_{ab} = e
(\partial_a A^c_\mu) \phi_{bc} + e (\partial_b A^c_\mu) \phi_{ac} +e
A^c_\mu \partial_c \phi_{ab} . \eqno ( 10 ) $$. It is important to
emphasize that all the dynamics is encoded entirely in the field $ A_\mu$.
The fields $\gamma_{\mu \nu }, \phi_{ab}$ act as backgrounds. Integrating
afterwards the Lagrangian w.r.t the internal $y$ variables yields an
effective $5$-dim spacetime theory with an $infinite$ number of
field-components. Similar actions have been constructed in full detail in
[16] related to spacetime gauge theories of the Virasoro and $w_\infty$
algebras, area-preseriving diffs of a plane. The $w_{1+ \infty}$ are
area-preserving diffs of a cylinder and the $ su(\infty) $ are
area-preserving diffs for the sphere. For relevant work on self dual
gravity of the cotangent bundle of a two-dim surafce and $w_\infty$
algebras see [10,13,15,18,21]. Other important work on $ W$ algebras can
be found in [5,11,12,17,18,19].  To see the role of the large $N$ limit
and the Moyal deformation quantization in the direct connection among
p-brane actions and (Generalized) Yang-Mills theories see [14]. In
particular, the connection between $W$-geometry and Fedosov's deformation
quantization was performed in [8,9]. In [10] it was shown how non-critical
$ w_\infty$ strings are effective $ 3d$ field theories that are devoid of
BRST anomalies in $ D = 27 $ for the bosonic case, and $ D = 11 $ for the
supersymmetric case. Noncritical $ w_\infty$ (super) strings behave like
critical (super) membranes. It is warranted to study the relation of $
w_\infty$ strings ( higher conformal spin field theories ) propagating on
curved $ AdS_4 \times S^7 $ backgrounds to higher spin field theories on $
AdS_4$ spaces (target space symmetry) [6,7]. Since the conformal boundary
of $ AdS_4$ is $ 3$-dimensional and the $ w_\infty$ strings are effective
$ 3$-dim theories, it is very natural to assume that $ w_\infty$ strings
actually live on the boundary of $ AdS_4 $. This result for $ m = n = 5 $
can be generalized for other values of $m$ and $n$. Since the scalar
curvature depends explicitly on the size of the throat (internal radius)
and the dimensions, one will have to tune these values in order to produce
an exact cancellation of the sum of the scalar curvatures.  Concluding,
higher dimensional pure gravitational actions restricted to $ AdS_m \times
S^n $ backgrounds admit a holographic reduction to a $lower$ $
m$-dimensional Yang-Mills-like gauge theory of diffs of $ S^n $
interacting with a charged non-linear sigma model plus boundary terms by a
simple tuning of the throat and internal radius sizes. After performing a
harmonic expansion of the fields with respect to the internal coordinates
of $S^n $ and a subsequent integration one obtains an $m$-dimensional
effective action involving an infinite-component field theory.  The
supersymmetrization program can be carried out in a straighforward.
fashion.
\bigskip
\centerline{\bf Acknowledgements}
We are indebted to
Prof. J. Mahecha and Dr. Alfred Schoeller for their valuable help and to
the Kuhlmann family for their kind hospitality in Heidelberg, Germany
where this work was completed.
\bigskip \bigskip
\centerline{\bf References}
\bigskip

1-J. Maldacena: Adv. Theor. Math. Phys {\bf 2} (1998) 231. hep-th/9711200.

2-J.L Petersen: ``Introduction to the Maldacena Conjecture''
hep-th/9902131.

3-Y.Cho, K, Soh, Q. Park, J. Yoon: Phys. Lets {\bf B 286} (1992) 251.: J.
H. Yoon: ``4-dimensional Kaluza-Klein approach to General Relativity in
the (2,2) splitting of spacetimes'' gr-qc/9611050 J. H. Yoon:
``Algebraically special class of spacetimes and (1,1) dimensional field
theories'' hep-th/9211129.

4- J. Hoppe: ``Quantum theory of a Relativistic Surface'' MIT Ph.D Thesis
1982.

5- P. Bouwknegt, K. Schouetens: ``$W$-symmetry in Conformal Field Theory''
Phys. Reports {\bf 223} (1993) 183-276.

6- M. Vasiliev: ``Higher Spin Gauge Theories, Star Product and AdS
spaces'' hep-th/9910096 M. Vasiliev, S. Prokushkin: ``$3D$ Higher-Spin
Gauge Theories with Matter''. hep-th/9812242, hep-th/9806236.

7-E Sezgin, P. Sundell: ``Higher Spin $N=8$ Supergravity in $AdS_4$''.
hep-th/9805125; hep-th/9903020.

8- B. Fedosov: Jou. Diff. Geometry {\bf 40} (1994) 213.

9-C. Castro: ``W-Geometry from Fedosov Deformation Quantization'' Jour.
Geometry and Physics {\bf 33} (2000) 173.

10-C. Castro: Jour. Chaos, Solitons and Fractals {\bf 7} (5) (1996) 711.
C. Castro: ``Jour. Math. Phys. {\bf 34} (1993) 681. C. Castro: Phys Lets
{\bf B 288 } (1992) 291.

11-E Nissimov, S. Pacheva, I. Vaysburd: ``$W_\infty$ Gravity, a Geometric
Approach''. hep-th/9207048.

12- G. Chapline: Jour. Chaos, Solitons and Fractals {\bf 10} (2-3) (1999)
311. Mod. Phys. Lett {\bf A 7} (1992) 1959. Mod. Phys. Lett {\bf A 5}
(1990) 2165.

13- H. Garcia-Compean, J. Plebanski, M. Przanowski:.hep-th/9702046.
``Geometry associated with SDYM and chiral approaches to Self Dual Gravity
``.

14-C. Castro: ``Branes from Moyal Deformation quantization of GYM''
hep-th/9908115. S. Ansoldi, C. Castro, E. Spallucci: Phys. Lett {\bf B
504} (2001) 174. S. Ansoldi, C. Castro, E. Spallucci: Class. Quant. Grav.
{\bf 18} (2001) L S. Ansoldi, C. Castro, E. Spallucci: Class. Quan.
Gravity {\bf 18} (2001) 2865. hep-th/0106028

15-Q.H. Park: Int. Jour. Mod. Phys {\bf A 7} (1992) 1415.

16-W. Zhao: Jour. Math. Phys. {\bf 40} (1999) 4325. Y. Cho, W. Zoh: Phys.
Review {\bf D 46} (1992) 3483.

17- E. Sezgin: ``Area-preserving Diffs, $w_\infty$ algebras, $w_\infty$
gravity'' hep-th/9202080. E. Sezgin: ``Aspects of $ W_\infty$ Symmetry''
hep-th/9112025.

18- C. Hull: Phys. Letts {\bf B 269 } (1991) 257.

19-Rodgers: ``BF ....'' hep-th/9203067

20- G. Dito, M. Flato, D. Sternheimer, L. Takhtajan: ``The deformation
quantization of Nambu-Poisson mechanics'' hep-th/9602016 H. Awata, M. Li,
G. Minic, T. Yoneya: ``Quantizing the Nambu Poisson brackets''
hep-th/9906248. G. Minic: ``M theory and Deformation Quantization''
hep-th/9909022

21- G. Chapline, K. Yamagishi: Class. Quan. Gravity {\bf 8 } (1991) 427.

22- C. S. Aulakh, D. Sahdev: Phy. Lets {\bf B 173 } (1986) 241.

\bye